\documentclass[twocolumn,showpacs,preprintnumbers,prc,superscriptaddress]{revtex4}
\usepackage{epsfig}
\usepackage{graphicx}
\usepackage{epstopdf}
\usepackage{amsmath,amssymb,amsfonts}
\usepackage{array}
\usepackage{url}
\usepackage{hyperref}
\usepackage{csquotes}
\usepackage{xspace}
\usepackage{textcomp}
\usepackage[usenames,dvipsnames]{color}
\newcommand{\sqsn}{\mbox{$\sqrt{s_{_{NN}}}$}\xspace}
\newcommand{\bef}{\begin{figure}}
\newcommand{\eef}{\end{figure}}
\newcommand{\bc}{\begin{center}}
\newcommand{\ec}{\end{center}}

\begin{document}
\title{Conserved charge fluctuations using the $D$-measure in heavy-ion collisions}

\author{D. K. Mishra}
\email {dkmishra@rcf.rhic.bnl.gov, dkmishra@barc.gov.in}
\affiliation{Nuclear Physics Division, Bhabha Atomic Research Center, 
Mumbai 400085, India}
\author{P.~K.~Netrakanti}
\affiliation{Nuclear Physics Division, Bhabha Atomic Research Center, 
Mumbai 400085, India}
\author{P.~Garg}
\affiliation{Department of Physics and Astronomy, Stony Brook University, SUNY, 
Stony Brook, New York 11794-3800, USA}

\begin{abstract}
We study the net-charge fluctuations, $D$-measure variable, in high energy 
heavy-ion collisions in heavy-ion jet interaction generator (HIJING), 
ultrarelativistic quantum molecular dynamics (UrQMD) and hadron resonance gas (HRG) 
models for various center-of-mass energies (\sqsn). The effect of kinematic 
acceptance and resonance decay, in the pseudorapidity acceptance interval 
($\Delta\eta$) and lower transverse momentum ($p_{T}^{min}$) threshold, on 
fluctuation measures are discussed. A strong dependence of $D$ with the $\Delta\eta$ 
in HIJING and UrQMD models is observed as opposed to results obtained from the HRG 
model. The dissipation of fluctuation signal is estimated by fitting the $D$-measure 
as a function of the $\Delta\eta$. An extrapolated function for  higher $\Delta\eta$ 
values at lower \sqsn is different from the results obtained from models. 
Particle species dependence of $D$ and the effect of the $p_T^{min}$ selection 
threshold are discussed in HIJING and HRG models. The comparison of $D$, at 
midrapidity, of net-charge fluctuations at various \sqsn obtained from the 
models with the data from the ALICE experiment is discussed. The results from the
present work as a function of $\Delta\eta$ and \sqsn will provide a baseline for 
comparison to experimental measurements.

\end{abstract}

\maketitle

\section{Introduction}
\label{intro} 
One of the major goals of the heavy-ion collisions is to study the phase 
structure of the quantum chromodynamic (QCD) phase diagram at finite 
temperature ($T$) and baryon chemical potential 
($\mu_B$)~\cite{Stephanov:1998dy}. Several studies suggest that, at low $T$ and 
large $\mu_B$, a first-order phase transition occurs from quark-gluon-plasma 
(QGP) to the hadronic phase. At high $T$ and low $\mu_B$, there is a cross over 
phase transition from QGP to the  hadron 
phase~\cite{Alford:1997zt,Stephanov:1996ki,Aoki:2006we,Pisarski:1983ms,
Ejiri:2008xt}. The study of event-by-event fluctuations provides a unique 
opportunity to describe the thermodynamic properties of the system created in 
heavy-ion collisions~\cite{Stephanov:1999zu,Asakawa:2000wh,Jeon:1999gr,
Ejiri:2005wq, Bazavov:2012vg}.

Event-by-event fluctuations of conserved quantities such as net-baryon number, 
net-electric charge and net-strangeness are proposed as a possible signal of 
the QGP formation and quark-hadron phase 
transition~\cite{Jeon:1999gr,Asakawa:2000wh}. Earlier studies suggest that, 
enhanced multiplicity fluctuations are connected to the production of QGP 
droplets, and suppression of fluctuations may occur due to the large 
difference in the degrees of freedom between the QGP and hadron gas 
phases~\cite{Asakawa:2000wh,Jeon:1999gr,Adams:2003st}. In the QGP phase, quarks with 
fractional unit charge 1/3 are the charge carriers, while in the hadronic phase 
hadrons are the charge carriers with unit charges. Hence, net-charge 
fluctuations in the QGP phase are significantly smaller as compared to those of the
hadron phase. These differences may be exploited as indicators of the 
formation of quark-gluon plasma in high-energy heavy-ion collisions. Thus, the 
net-charge fluctuations are strongly dependent on the phase of their origin. Due to 
the rapid expansion of the fireball created in the heavy-ion 
collisions, the fluctuations created in the initial state may survive until the 
freeze-out~\cite{Asakawa:2000wh}. The fluctuations of the net-charge depend on 
the square of the charge states present in the system. The net-charge 
fluctuation is plagued by uncertainties due to volume fluctuation, resonance 
decay, exact local charge conservation, or repulsive forces among 
hadrons~\cite{Alba:2014eba}. Experimentally, only a fraction of the particles 
are measured in the detector acceptance, which may be subject to charge 
conservation effects. Also the measured fluctuations depend on the survival 
probability of the charge fluctuations during the hadronization process.
The conservation laws limit the dissipation of the fluctuations which 
suffer after the hadronization has occurred. This dissipation occurs by diffusion. 
The heavy-ion collisions forming QGP that hadronizes at a time $\tau_0$, produces 
anomalous dynamic charge fluctuations~\cite{Aziz:2004qu}. The hadronic diffusion 
from $\tau_0$ to a freeze-out time $\tau_f$ can dissipate these fluctuations. It is 
observed that, there is a decrease in dynamical charge fluctuation as a function of 
$\tau_f$. Due to the diffusion of particles in rapidity space, these fluctuations 
may also get diluted in the expanding medium~\cite{Shuryak:2000pd,Aziz:2004qu}. It 
is argued that, the reduction of the fluctuation in the QGP phase might be observed 
only if the fluctuations are measured over a large rapidity 
range~\cite{Shuryak:2000pd}. The QGP suppression of the charge fluctuation is not 
observed in the experimental data, however the data are consistent with the 
diffusion estimates. The effect of the critical fluctuations is crucially different 
from the QGP suppression. While the QGP suppression is the history effect, the 
critical fluctuations are the equilibrium fluctuations pertaining to the freeze-out 
point, and the diffusion is necessary to establish them~\cite{Hatta:2003wn}.

In the heavy-ion experiments, the collision volume is not directly measured, 
hence one can get rid of the volume, to first order by taking the ratios of the 
number of positive ($N_+$) and negative ($N_-$) particles. The variance of the 
ratio of positive and negative particles scaled by the total number of charged 
particles is defined as the $D$-measure ($D$), of the net-charge, which provides a 
measure of charge fluctuations per unit entropy. The $D$ is related to 
the ratio 
$R$ $(= N_+/N_-)$ as~\cite{Jeon:2003gk,Pruneau:2002yf}:
\begin{eqnarray}
 D =  \langle N_{\mathrm{ch}}\rangle\langle \delta R^2\rangle &=& 
\frac{4}{\langle 
N_{\mathrm{ch}}\rangle}\langle \delta N_+^2 + \delta N_-^2 - 2\delta N_+ \delta 
N_-\rangle \nonumber \\
&\approx& \frac{4\langle \delta Q^2\rangle}{\langle N_{\mathrm{ch}}\rangle}
\label{eq:Dmea}
\end{eqnarray}
where $\langle \delta Q^2 \rangle$ is the variance of the net-charge with $Q = 
N_+ - N_-$ being the difference between $+$ve and $-$ve particles and 
$N_{\mathrm{ch}} = N_+ + N_-$ being the total number of charged particles 
measured in a particular phase-space. Assuming the negligible quark-quark 
interactions, the $D$ is found to be approximately four times smaller in the QGP 
phase as compared to the hadron gas phase~\cite{Jeon:2003gk}. For uncorrelated pion 
gas, $D_{\pi}$ is estimated to be 4 and by taking the resonance contributions the 
value reduces to 3. For noninteracting massless quarks and gluons $D_{QGP}$ is 
found to be a factor of 5 smaller than the $D_{\pi}$. In the constituent quark 
scenario, it is reported that the $D$-measure value might be 
$\sim3.3$~\cite{Bialas:2002hm}. The $D$-measure results from 
ALICE~\cite{Abelev:2012pv} 
seems to suggest that the hadronization may be due to coalescence which is not the 
case here, although the fluctuations suggest the existence of non-interacting 
equilibrated-QGP before the freeze-out. Hence, the measurement of $D$ can be a 
useful 
observable to distinguish between the QGP and hadron gas phase~\cite{Jeon:1999gr}. In 
Ref.~\cite{Bleicher:2000ek}, it is reported that, there is no significant difference 
found for $D$ values at SPS~\cite{Sako:2004pw} and RHIC~\cite{Abelev:2008jg} 
energies. The $D$ value at \sqsn = 200 GeV is observed to be close to the hadron gas 
prediction~\cite{Abelev:2008jg}. However, recent results from LHC at \sqsn = 2.76 
TeV~\cite{Abelev:2012pv} show that the measured $D$ value is significantly lower than 
that measured at SPS and RHIC energies~\cite{Abelev:2008jg,Sako:2004pw}. The value 
of $D$ at LHC energy approaches the expectation from the QGP phase. 

Experimentally, the net-charge fluctuations are studied in terms of dynamical 
fluctuation measure $\nu_{(\pm,\mathrm{dyn})}$, which is found to be independent 
of detection efficiencies. The quantity $\nu_{(\pm,\mathrm{dyn})}$ is defined by;

\begin{equation}
 \nu_{(\pm,\mathrm{dyn})} = \frac{\langle N_+(N_+ - 1)\rangle}{\langle 
N_+\rangle^2} + \frac{\langle N_-(N_- - 1)\rangle}{\langle N_-\rangle^2} - 
2\frac{\langle N_-N_+\rangle}{\langle N_-\rangle \langle N_+ \rangle}
\label{eq:nudyn}
\end{equation}
where $\langle N_+\rangle$ and $\langle N_-\rangle$ are the average number of 
positive and negative charged particles measured within the detector acceptance.
The quantity $\nu_{(\pm,\mathrm{dyn})}$ is a measure of relative correlation of 
\enquote{$++$}, \enquote{$--$} and \enquote{$+-$} charged particle pairs. A 
positive value of $\nu_{(\pm,\mathrm{dyn})}$ implies the correlation of same 
charge pairs, where as a negative value indicates the contribution dominated by 
opposite charge pairs. The $D$ and $\nu_{(\pm,\mathrm{dyn})}$ are related to 
each other by~\cite{Jeon:2003gk};

\begin{equation}
 \langle N_{\mathrm{ch}}\rangle \nu_{(\pm,\mathrm{dyn})}\approx D - 4
\end{equation}
Keeping in mind the importance of the $D$ variable for the conserved 
number fluctuation, we have estimated the observable in the hadron resonance gas 
(HRG) model with realistic acceptance effects.

In this work, we demonstrate the effect of different pseudo-rapidity 
($|\Delta\eta|$) intervals and the lower transverse momentum ($p_T^{min}$) particle 
selection threshold on the $D$ in the framework of a HRG model and other heavy-ion 
models such as heavy-ion jet interaction generator (HIJING)~\cite{Wang:1991hta} and 
ultrarelativistic quantum molecular dynamics (UrQMD)~\cite{Bleicher:1999xi}. Further, 
we show the effect of resonance decay on the studied observable. It is important to 
find an adequate baseline for the conserved quantities particularly for net-charge 
fluctuations due to contributions from the higher charge states, resonance 
decays, and also contributions from quantum statistics for lower mass particles 
such as pions and kaons. Since net-charge fluctuations are dominated by 
contributions from pions, therefore measured fluctuations are strongly 
influenced by quantum statistics effects. The present study is an attempt to provide 
a more realistic baseline comparison to the corresponding experimentally 
measured quantities. 

The paper is organized as follows: In the following section, we discuss the HRG 
model used in this study as well as the implementation of resonance decays. We 
also briefly discuss the HIJING and UrQMD heavy-ion models which have been used for 
comparison with the results obtained from the HRG model. In Sec. \ref{sec:results}, 
the results for the $D$ observable for different energies, $|\Delta\eta|$ acceptance 
and effect of the $p_T^{min}$ selection threshold are discussed. The diffusion in 
rapidity space is studied at different collision energies and presented in the same 
section. Finally, in Sec.~\ref{sec:summary} we summarize our findings and mention the 
implication of this work.

\section{The $D$-measure from different model calculations}
\label{sec:hrg}
In this section, we discuss different models such as HRG, HIJING and 
UrQMD to understand the $D$ values measured in the experiment. These 
models provide the proper baseline to compare with the experimental data.

\subsection{Hadron resonance gas model}
The HRG model has been successfully applied to explain the particles produced 
in heavy-ion collisions from AGS, RHIC, and LHC 
energies~\cite{BraunMunzinger:2003zd,Cleymans:2005xv,Andronic:2011yq}. 
Further, the susceptibilities and their ratios in the hadronic phase calculated 
in the HRG model reasonably agree with the lattice QCD calculations at lower 
$\mu_B$ values~\cite{Bazavov:2012vg}. Several studies have been performed with 
the HRG model for the fluctuation of conserved quantities, which are considered 
as a baseline for such 
measurements~\cite{Begun:2006jf,Karsch:2010ck,Becattini:2005cc,Garg:2013ata,
Fu:2013gga,Rau:2013xya,Mishra:2016tne}. 

The partition function ($Z$) in the HRG model includes all relevant degrees of 
freedom of the confined, strongly interacting matter and contains all the 
interactions that result in resonance formation~\cite{Karsch:2010ck}. The heavy-ion 
experiments covering limited phase space can access only part of the total 
number of produced particles in the collisions. Hence, the grand canonical ensemble 
is more appropriate to describe such a scenario~\cite{Koch:2008ia}.
Assuming a thermal system produced in the heavy-ion collisions, the 
thermodynamic pressure ($P$) can be written as a sum of the partial pressures of 
all the considered particle species $i$ which can be baryon ($B$) or meson 
($M$) at temperature $T$ and chemical potential $\mu_B$:

\begin{equation}
 P(T,V,\mu_i) = \frac{T}{V}\mathrm{ln}Z_i    
\end{equation}
where $\mathrm{ln}Z_i=$ $\sum_M \mathrm{ln}Z_i (T,V,\mu_i)+\sum_B 
\mathrm{ln}Z_i (T,V,\mu_i)$ and $\mathrm{ln}Z_i$ of individual particles 
can be written as:

\begin{equation}
 \mathrm{ln} Z_i(T, V, \mu_i) = \pm \frac{Vg_i}{2\pi^2}\int d^3p 
~\mathrm{ln}\big\{1\pm \mathrm{exp}[(\mu_i-E)/T]\big\},
\end{equation}
where $V$ is the volume of the system, $g_i$ is the degeneracy factor of the 
$i$-th particle and the $\pm$ve signs are for baryons and mesons, respectively. 
The total chemical potential of the individual particle is $\mu_i$ = 
$B_i\mu_{B} + Q_i\mu_{Q} + S_i\mu_{S}$, where $B_i$, $Q_i$, and $S_i$ are the 
baryon, electric charge  and strangeness number of the $i$-th particle, with 
corresponding chemical potentials $\mu_{B}$, $\mu_{Q}$ and $\mu_{S}$, 
respectively. The collision energy dependence of freeze-out parameters ($\mu_B$ 
and $T_{\mathrm{f}}$) is used as given in Ref.~\cite{Cleymans:2005xv} and the 
parametrized $\mu_Q$ and $\mu_S$ are given in Ref.~\cite{Karsch:2010ck}. The 
volume element ($d^3p$) of a particle of mass $m$ in a static 
fireball can be written as $d^3p=p_Tm_T \mathrm{cosh}\eta dp_Td\eta d\phi$ 
and energy ($E = m_T \mathrm{cosh}\eta$) of the particle, where $m_T$ 
corresponds to the transverse mass of the particle ($\sqrt{m^2 + p_T^2}$) with 
$p_T$, $\eta$, and $\phi$ being the transverse momentum, pseudo-rapidity, and 
azimuthal angle, respectively. One can apply the acceptance cuts on these 
variables to compare with the experimental results. The $n$-th order generalized 
susceptibilities ($\chi$) are written as~\cite{Garg:2013ata}:

\begin{equation}
 \chi_i^{(n)} = \frac{d^n[P(T,\mu)/T^4]}{d(\mu_i/T)^n}.    
\end{equation}
For mesons, $\chi_i$ can be expressed as;
\begin{eqnarray}
 \chi_{i,\mathrm{meson}}^{(n)}=\frac{I^n}{VT^3}
\int{d^{3}p}\sum_{k=0}^{\infty}(k+1)^{n-1} \nonumber \\  
\times ~ \mathrm{exp}\bigg\{\frac {-(k+1)E } {T}\bigg\} {\mathrm{exp}\bigg\{ 
\frac{(k+1)\mu}{T}\bigg\}},
\label{eq:susc_mes}
 \end{eqnarray}
 and for baryons,
\begin{eqnarray}
\chi_{i,\mathrm{baryon}}^{(n)}=\frac{I^n}{VT^3} 
\int{d^{3}p}\sum_{k=0}^{\infty}{(-1)^k}
(k+1)^{n-1} \nonumber \\ 
\times ~ \mathrm{exp}\bigg\{\frac{-(k+1)E} {T}\bigg\} 
{\mathrm{exp}\bigg\{\frac{(k+1)\mu}{T}\bigg\}},
\label{eq:susc_bar}
\end{eqnarray}
where $I$ represents either $B_i$, $Q_i$ or $S_i$ of the $i$th particle 
depending on whether the susceptibility $\chi_{i}$ represents for net-baryon, 
net-electric charge, or net-strangeness. The total generalized susceptibilities 
will be the sum of susceptibility of mesons and baryons. 

The experimentally measured stable particles (pions, kaons, and protons along 
with their anti-particles) have contributions from primordial as well as from 
resonance decay. Neutral resonances introduce positive correlations between 
$N_+$ and $N_-$ and hence the decayed daughters from resonances can affect the 
fluctuation of the final measured particles. The generalized $n$-th order 
susceptibility for stable particle $i$ can be written 
as~\cite{Nahrgang:2014fza,Mishra:2016qyj};

\begin{equation}
\chi_i^{(n)} = \chi_i^{*(n)} + \sum_R \chi_R^{(n)}\langle n_i\rangle ^n_R
\label{eq:chi_ave}
\end{equation}
The first term in Eq.~\ref{eq:chi_ave} corresponds to the contribution from 
primordial yield and the second term corresponds to the contribution from the 
resonance particles to stable particles. The summation runs over all the 
resonance states which contribute to the final particle $i$ and $\langle n_i 
\rangle_R = \sum_r b_r^R n_{i,r}^R$ is the average number of particle type $i$ 
produced from the resonance $R$. Further, $b_r^R$ is the branching ratio of the 
$r$-th decay channel of the resonance $R$ and $n_{i,r}^R$ is the number of 
particle $i$ produced in that decay branch. In this study, we have considered 
the fluctuation in the resonance production and the fluctuation in the produced 
daughters from the resonance. Detailed discussion on resonance decay can be 
found in Refs.~\cite{Nahrgang:2014fza,Mishra:2016qyj}. 

\subsection{D-measure using HIJING and UrQMD model}
In the present study, we discuss the $D$ observable calculated using 
different heavy-ion models. Here we briefly discuss event generators such 
as HIJING~\cite{Wang:1991hta} and UrQMD models~\cite{Bleicher:1999xi}. 
HIJING is a perturbative QCD model which produces minijet partons that are 
later transformed into string fragments that then fragment into hadrons. The 
cross sections for hard parton scattering are calculated using the leading order 
in order to account for the higher-order corrections, and a $K$-factor is invoked. 
The diquark-quark strings with gluon kinks induced by soft gluon radiation are used 
to calculate the soft contributions. Jet quenching and shadowing can also be 
treated in this approach~\cite{Zhang:2002dy}. The UrQMD is a microscopic 
transport approach based on the propagation of constituent quarks and diquarks 
accompanied by mesonic and baryonic degrees of freedom~\cite{Bleicher:1999xi}. 
It simulates multiple interactions of baryon-baryon, meson-baryon, and 
meson-meson pairs. The model preserves the conservation of baryon number, 
electric charge and strangeness number. It also models the baryon-stopping 
phenomena which is one of the essential features in high-energy heavy-ion 
collisions particularly at lower collision energies. In this model, the 
space-time evolution of the fireball is studied in terms of excitation and 
fragmentation of color strings, and the formation and decay of hadronic 
resonances~\cite{Bleicher:2000ek,Sharma:2015lva}.

\section{Results and discussion}
\label{sec:results}
The measured fluctuations may get diluted during evolution of the system from 
hadronization to the kinetic freeze-out because of the diffusion of the charged 
particles~\cite{Aziz:2004qu}. It is proposed to study the net-charge 
fluctuations as a function of rapidity interval, which has been explored with 
the ALICE experiment at LHC by studying $D$ as a function of the pseudorapidity 
interval ($\Delta\eta$)~\cite{Abelev:2012pv}. 
\bef[h]
\bc
\includegraphics[width=0.5\textwidth]{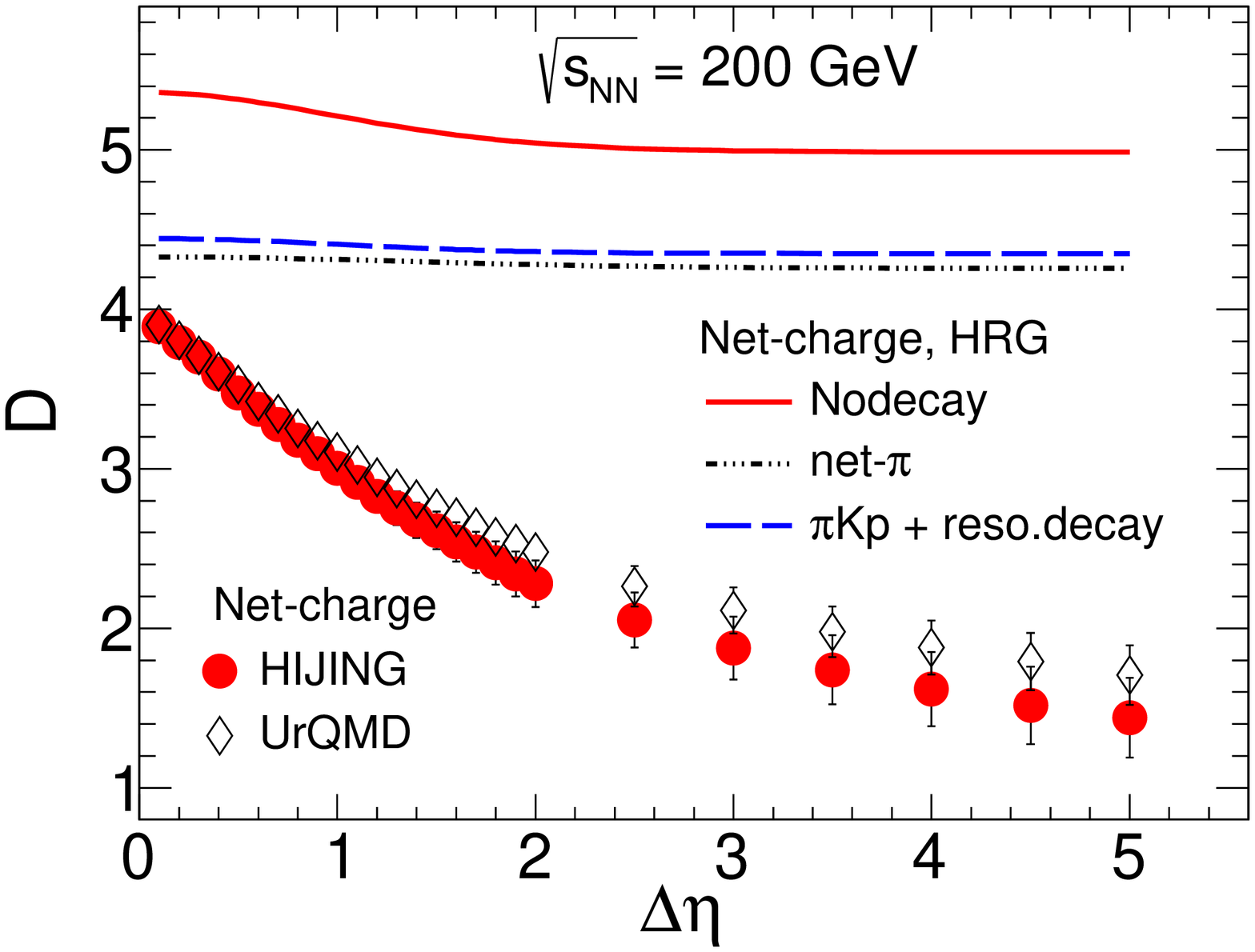}
\caption{Fluctuation, $D$, for net-charge as a function of 
$\Delta\eta$ for (0\%--5\%) centrality in Au$+$Au collisions at \sqsn 
= 200 GeV with HIJING and UrQMD models. The $D$ from HRG calculations 
for net-charge without resonance decay (solid line), with resonance decay 
(dashed line) and with net-pion (dashed-dotted line) are also shown.}
\label{fig:etaVsD_model_200gev}  
\ec
\eef

\bef[t]
\bc
\includegraphics[width=0.5\textwidth]{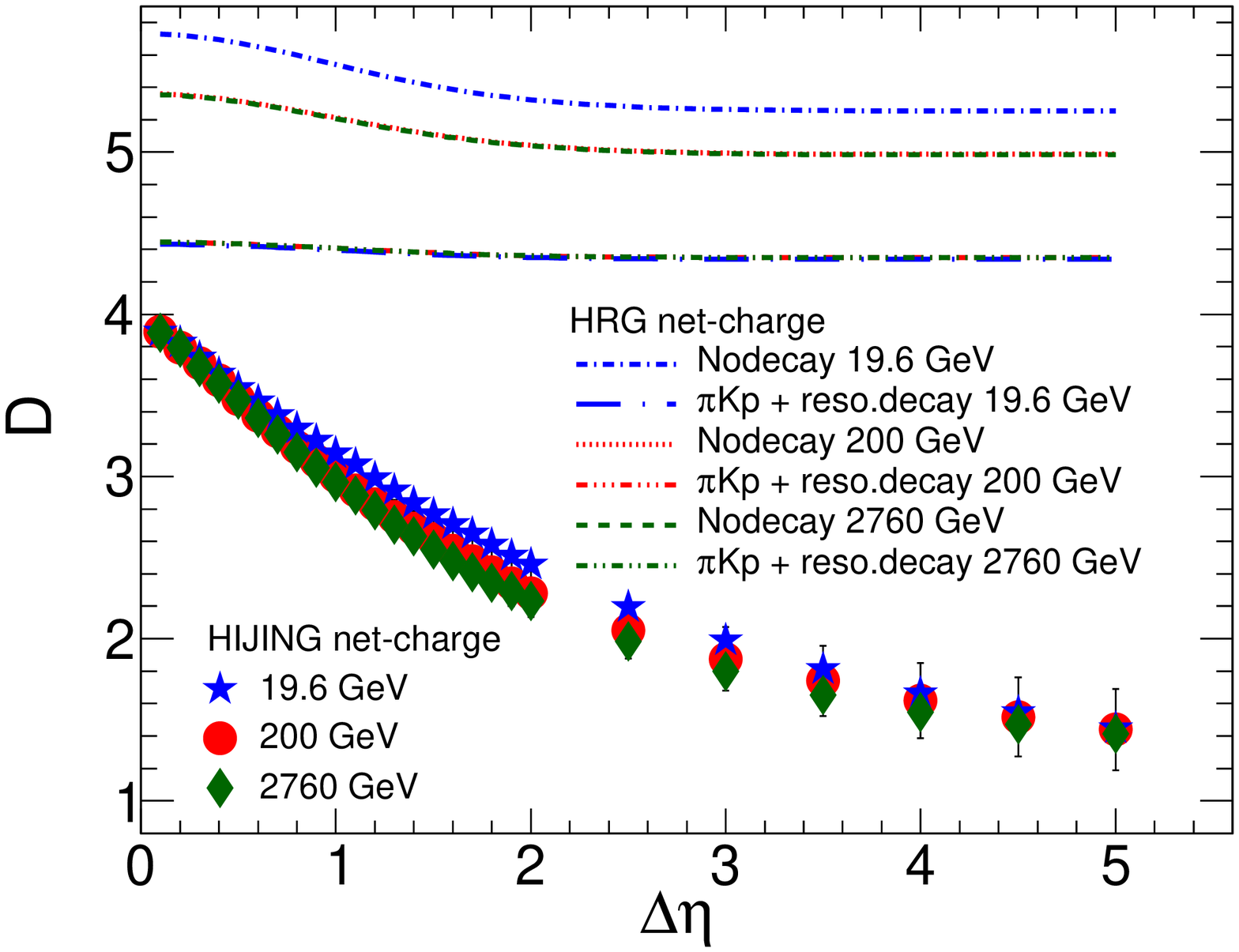}
\caption{The $D$ values for net-charge fluctuations as a function of 
$\Delta\eta$ for (0\%--5\%) centrality in Au$+$Au collisions at \sqsn 
= 19.6, 200 and 2760 GeV with the HIJING model. The $D$ from HRG, with and without 
resonance decay contribution, compared with model calculations at different 
\sqsn energies.}
\label{fig:etaVsD_ene_model} 
\ec
\eef

\begin{figure*}[ht]
\bc
\includegraphics[width=1.0\textwidth]{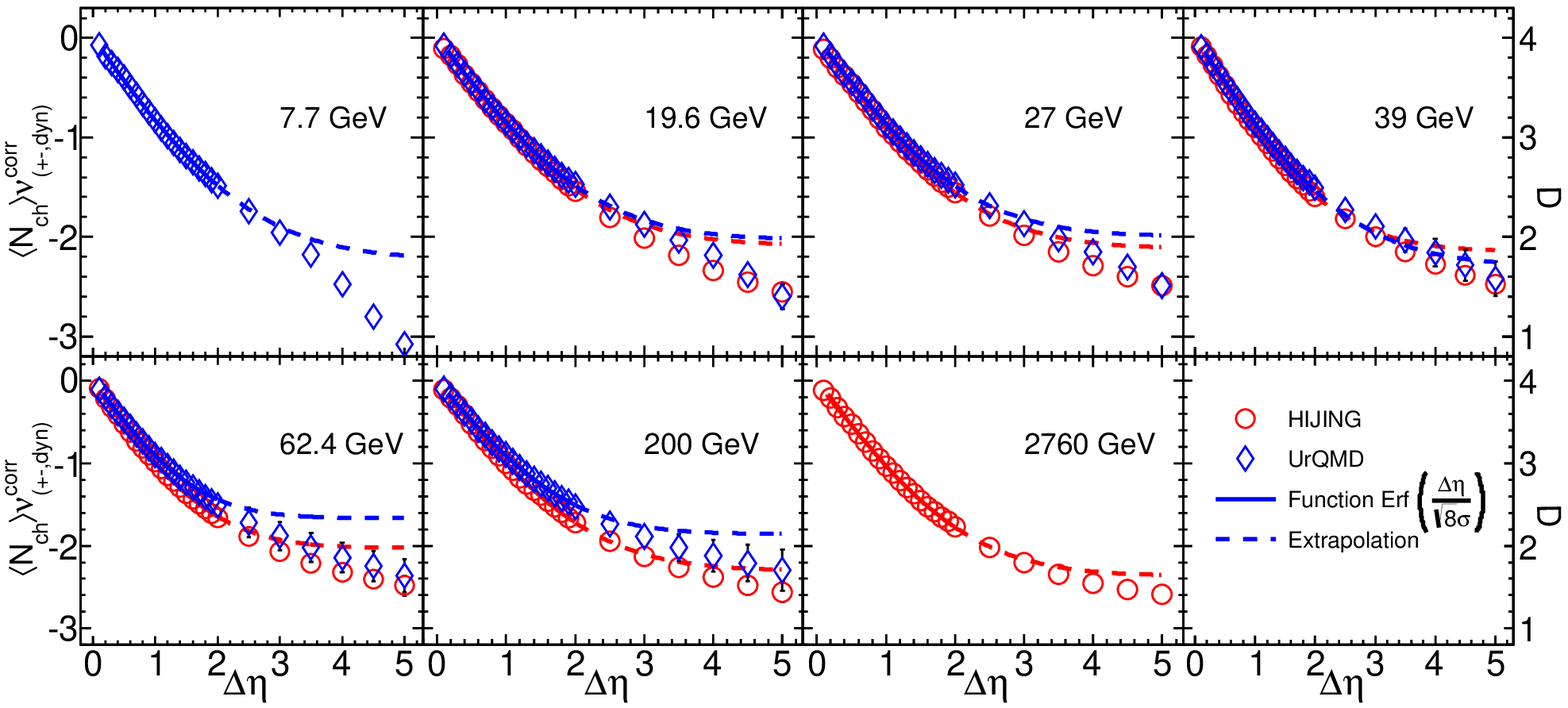}
\caption{The $\langle N_{ch}\rangle \nu_{(\pm,\mathrm{dyn})}^{corr}$ (left 
axis) and $D$ (right axis) as a function of $\Delta\eta$ intervals for 
(0\%--5\%) centrality in Au$+$Au collisions at different \sqsn. The data points 
are fitted with the $Erf(\Delta\eta/\sqrt{8}\sigma)$ up to $\Delta\eta$ = 2.0 
shown in solid line and extrapolation up to higher $\Delta\eta$ intervals is 
shown in dashed line.}
\label{fig:etaVsD_fit}
\ec
\end{figure*}

It is reported that, the $D$-measure has a decreasing trend with a 
flattening tendency at larger $\Delta\eta$ values~\cite{Abelev:2012pv}. 

\subsection{$D$-measure from different models}
We have investigated this study with HIJING and UrQMD event generators up to large 
$\Delta\eta$ interval at different \sqsn. As discussed in the previous section, 
although both the models are based on different physics inputs, these models have 
been successfully applied to explain the experimental data at RHIC energies. 
Figure~\ref{fig:etaVsD_model_200gev} shows the $D$ as a function of $\Delta\eta$ 
interval. All the stable charged particles having $p_T$ within 0.2 to 5.0 
GeV/$c$ are considered for the analysis. The $D$ values from both the models 
agree well for all the $\Delta\eta$ intervals within the statistical 
uncertainties. The $D$ value decreases as a function of $\Delta\eta$. The higher 
$D$ value at smaller $\Delta\eta$ interval suggests that the correlation is 
maximum for the smaller $\Delta\eta$ interval which gets diluted at larger 
intervals. The curvature of $D$ shows a decreasing slope up to higher 
$\Delta\eta$ interval. This is in contrast to the observation made by ALICE 
experiment at \sqsn = 2.76 TeV, which predicts a flattening trend at large 
$\Delta\eta$ by extrapolation of the fitted curve to the higher $\Delta\eta$ 
region~\cite{Abelev:2012pv}. The $D$ values from the HRG calculation are 
compared with the HIJING and UrQMD results. The HRG calculation for net-charge 
fluctuations is performed within the same kinematic acceptance as done for 
other models. All the charged hadrons of mass up to 2.5 GeV as listed in the 
particle data book are considered. The HRG calculations for net-charge 
fluctuations are performed by considering all charged particles without 
resonance decay, only primordial pions, and stable charged particles ($\pi$, 
$K$, and $p$) with resonance decay contributions. Unlike HIJING and UrQMD model 
results, the HRG calculations do not show $\Delta\eta$ dependence of $D$. In 
case of no decay of resonances, there is small dependence observed at lower 
$\Delta\eta$ intervals. However, there is substantial decrease of $D$ value with 
inclusion of resonance decay contributions as compared to without decay of 
resonances. Since charge fluctuation is dominated by fluctuations from the 
pions, we have also compared the $D$ values for primordial pions by taking 
quantum statistics into account. The $D$ values for net-pions are closer to the 
results obtained from net-charge fluctuation. The calculation of $D$ from the HRG 
model will provide a pure thermal baseline as a function $\Delta\eta$. 

\subsection{Energy dependence of $D$-measure}
The $\Delta\eta$ dependences of the $D$ observable obtained from both HIJING and 
UrQMD models are very similar. Hence for the energy dependence studies, we 
consider results from the HIJING model. In order to study the energy 
dependence of $D$ as a function of $\Delta\eta$ intervals, we have considered 
simulated events for three different energies \sqsn = 19.6, 200 and 2760 GeV 
using HIJING. Figure~\ref{fig:etaVsD_ene_model} shows the $\Delta\eta$ 
dependence of $D$ of net-charge fluctuations for different energies. The $D$ 
values are consistently decreasing for all the studied energies up to large 
$\Delta\eta$ interval. We do not observe flattening behavior at higher 
$\Delta\eta$ intervals for any of the considered energies as predicted by the 
extrapolation of experimental data~\cite{Abelev:2012pv}. The $D$ values 
obtained from the HRG calculations are compared with results from HIJING. In 
case of without decay of resonances, the HRG result shows a higher value for 19.6 
GeV than the other two higher energies and remains constant as a function 
$\Delta\eta$ intervals, except small dependence at lower $\Delta\eta$ intervals. 
However, the $D$ values agrees well for all studied energies with inclusion of 
resonance decay contributions in the HRG calculations.

\subsection{Extraction of diffusion parameter}
\begin{figure}[ht]
\bc
\includegraphics[width=0.5\textwidth]{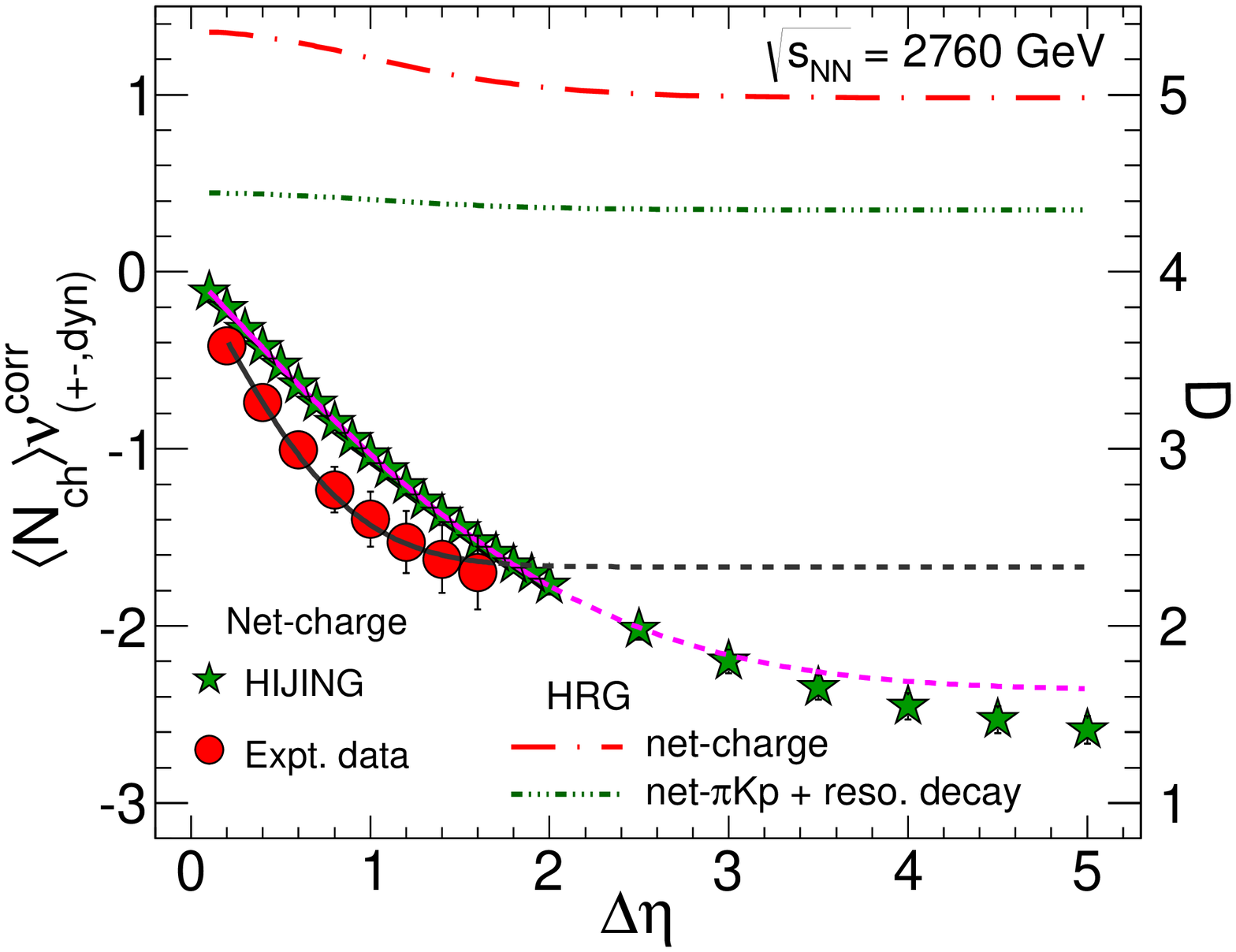}
\caption{The $\langle N_{ch}\rangle \nu_{(\pm,\mathrm{dyn})}^{corr}$ (left 
axis) and $D$ (right axis) as a function of $\Delta\eta$ intervals for 
(0\%--5\%) centrality in Pb$+$Pb collisions at different \sqsn. The data points 
are fitted with the $Erf(\Delta\eta/\sqrt{8}\sigma)$ up to $\Delta\eta$ = 2.0 
shown in solid line and extrapolation up to higher $\Delta\eta$ intervals is 
shown in dashed line. The $D$ from HRG, with and without resonance decay 
contribution, compared with the experimental data and HIJING model calculation.}
\label{fig:etaVsD_lhc}
\ec
\end{figure}

The measured fluctuation signals may get obliterated during the evolution of 
the system in heavy-ion nuclear collisions because of the diffusion of the 
charged particles in rapidity space. It is estimated that, the net-charge 
fluctuations induced by quark gluon plasma hadronization may survive diffusion 
in the hadronic stage~\cite{Aziz:2004qu}. It is also discussed  how 
much the fluctuations are reduced with the increase of accepted rapidity 
interval. It is observed from the experimental data at \sqsn = 2.76 TeV that the 
$D$ value has a decreasing slope at lower $\Delta\eta$ intervals and gets 
flattened at higher $\Delta\eta$ intervals~\cite{Abelev:2012pv}. 
Figure~\ref{fig:etaVsD_fit} shows the 
$\langle N_{\mathrm{ch}}\rangle\nu_{(\pm,\mathrm{dyn})}^{corr}$ and $D$ as a 
function of $\Delta\eta$ interval for (0\%--5\%) centrality in Au$+$Au 
collisions at different \sqsn using HIJING and UrQMD models. In a similar way as 
in Ref.~\cite{Abelev:2012pv}, the simulated data points are fitted with the 
error function, $Erf(\Delta\eta/\sqrt{8}\sigma)$ representing the diffusion in 
rapidity space~\cite{Aziz:2004qu}. This accounts for the broadening of the 
rapidity distributions due to interactions and particle production. The quantity 
$\sigma$ which characterizes the diffusion at freeze-out is obtained by fitting 
the $\langle N_{\mathrm{ch}}\rangle\nu_{(\pm,\mathrm{dyn})}^{corr}$ values up 
to $\Delta\eta=$ 2.0.
The results from both HIJING and UrQMD models do not show flattening behavior for 
$\Delta\eta$ above 1.5. The $\langle 
N_{\mathrm{ch}}\rangle\nu_{(\pm,\mathrm{dyn})}^{corr}$ and $D$ values from the models 
keep on decreasing even at higher $\Delta\eta$ intervals. 
Figure~\ref{fig:etaVsD_lhc} shows the comparison of $\langle 
N_{\mathrm{ch}}\rangle\nu_{(\pm,\mathrm{dyn})}^{corr}$ and $D$ as a 
function of $\Delta\eta$ obtained from the experimental 
data in ~Ref.~\cite{Abelev:2012pv} and the model calculations at \sqsn = 2760 GeV. 
The experimental data show flattening behavior of the fluctuations at higher 
$\Delta\eta$ as compared to the HIJING model calculations. The slope of the 
experimental data and the HIJING model calculations are different, which results in 
different extrapolated values at higher $\Delta\eta$. With larger uncertainties in 
the experimental data, it may so happen that the extrapolated values at higher 
$\Delta\eta$ will follow the experimental measurements. Further, the HRG calculations 
do not show $\Delta\eta$ dependence of $D$ in both without and with resonance decay 
contributions. The experimental data~\cite{Abelev:2012pv} and the present study with 
HIJING and UrQMD models follow the diffusion trend.

\begin{figure}[h]
\bc
\includegraphics[width=0.5\textwidth]{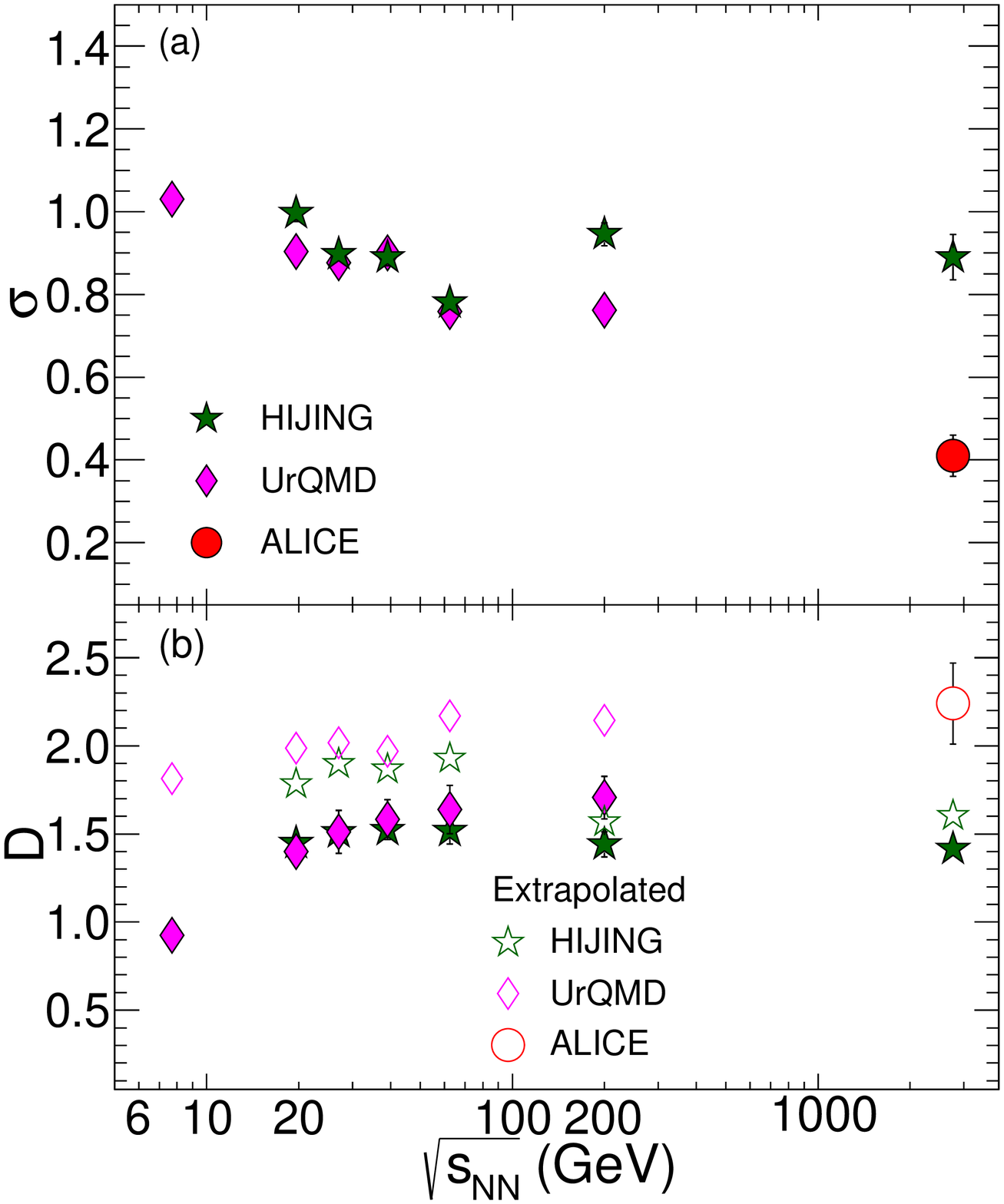}
\caption{Collision energy dependence of $\sigma$ (upper panel) which 
characterizes the diffusion at the freeze-out are calculated using HIJING and 
UrQMD models for (0\%--5\%) centrality in Au$+$Au collisions. The $D$ 
values as a function of \sqsn are shown in the lower panel. The solid symbols are 
values extracted from the simulated data at $\Delta\eta$ = 5.0 and the open 
symbols are extrapolated values of $D$ using the functional form at the same 
$\Delta\eta$ interval.}
\label{fig:diff_coeff_ene}
\ec
\end{figure}

The extrapolation of the fitted curves in Fig.~\ref{fig:etaVsD_fit} do not explain 
the $D$ values extracted at higher $\Delta\eta$ intervals particularly at lower 
energies. Hence, in order to estimate the diffusion parameters from $D$ values, it 
is important to measure up to higher $\Delta\eta$ intervals. The resulting values of 
$\sigma$ obtained from the above fit function at different \sqsn are shown in 
Fig.~\ref{fig:diff_coeff_ene}a. The collision energy dependence of the $D$ 
values is shown in Fig~\ref{fig:diff_coeff_ene}b. The $D$ values shown in 
Fig~\ref{fig:diff_coeff_ene}b are extracted from two different methods. The 
values extracted from the simulated data at $\Delta\eta$ = 5.0 are shown in 
solid symbols and the $D$ values calculated from the extrapolation of the fitted 
curve up to higher $\Delta\eta$ at 5.0 are shown in open symbols. The $D$ values 
estimated using the extrapolation method show higher values as compared to the 
values from the data points at $\Delta\eta$ = 5.0 obtained from the models. This 
can also be observed in Fig.~\ref{fig:etaVsD_fit} at higher $\Delta\eta$ 
intervals. However, the data points from the HIJING simulation are in better 
agreement with the extrapolated curve for \sqsn = 2.76 TeV. 

\bef[ht]
\includegraphics[width=0.5\textwidth]{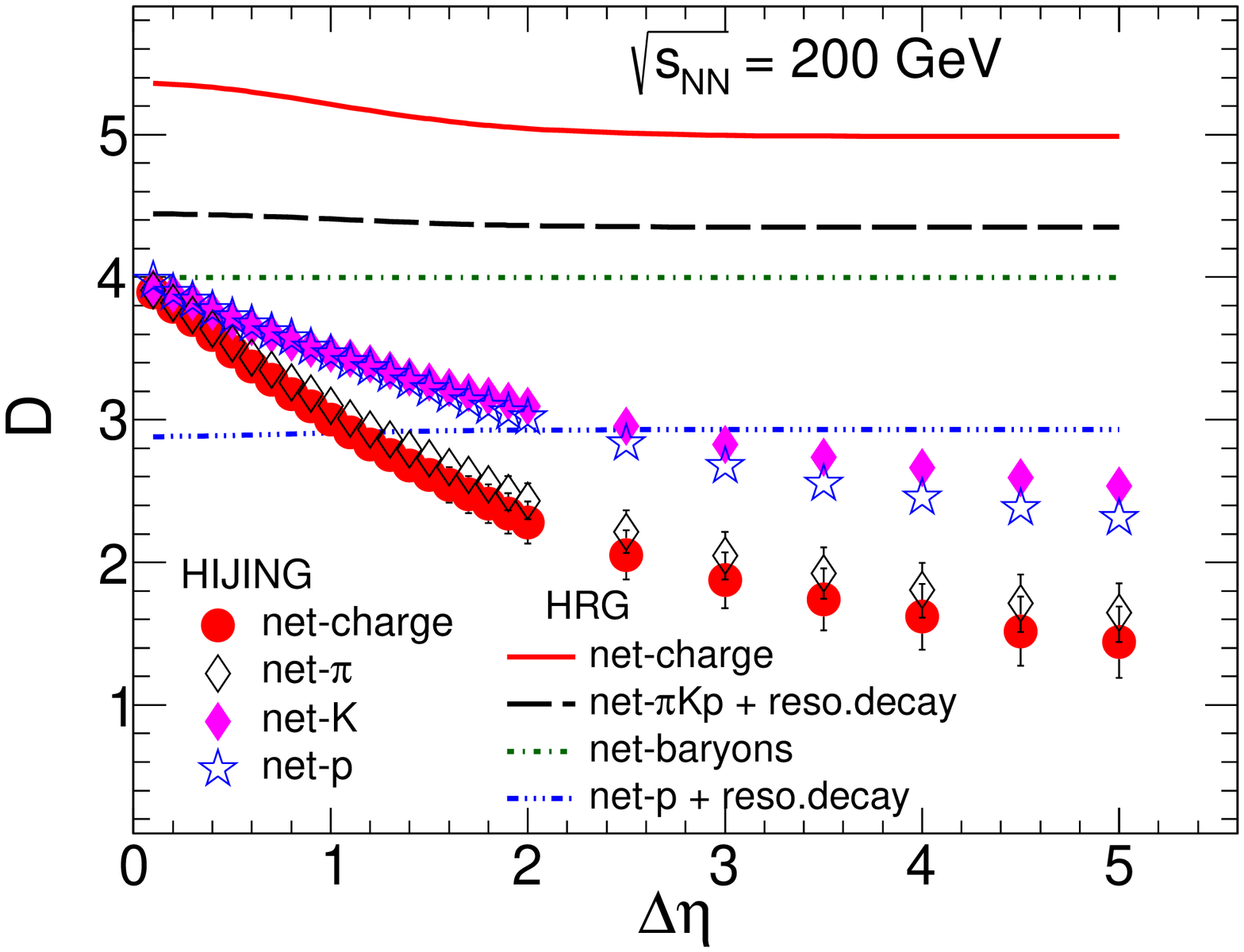}
\caption{The $D$ values for net-charge, net-pion, net-kaon, and net-proton 
fluctuations as a function of $\Delta\eta$ for (0\%--5\%) centrality in Au$+$Au 
collisions at \sqsn = 200 GeV with the HIJING model. The $D$ from HRG 
calculations for net-charge and net-baryon without and with inclusion of 
resonance decay are compared with model calculations.}
\label{fig:etaVsD_part_200gev} 
\eef

\subsection{Particle dependence of $D$-measure}
The net-charge fluctuation is mostly dominated by the fluctuation contribution
from pions. Hence it is important to identify the individual contributions of 
the stable charged hadrons to the net-charge fluctuation. 
Figure~\ref{fig:etaVsD_part_200gev} shows the $D$ of net-charge fluctuations as 
a function of $\Delta\eta$ interval in Au$+$Au collisions at \sqsn = 200 GeV 
from the HIJING model. The $D$ for net-pions ($\pi^+ - \pi^-$), net-kaons ($K^+ - 
K^-$), and net-protons ($p - \bar p$) are also shown. In the HIJING model we also 
observe that the $D$ values for net-charge fluctuations are dominated by net-pion 
fluctuations. The $D$ values of individual net-protons and net-kaons are similar 
and decrease with increasing $\Delta\eta$ interval. The studied observables for 
net-charge and net-baryon as a function of $\Delta\eta$ interval calculated in 
the HRG model are also compared. The net-charge and net-baryon values are 
compared for calculations with and without resonance decay contributions. Due to the 
contributions from the resonance decay, the $D$ values are substantially smaller 
when compared to calculations taking all the charged particles or all the 
baryons. Unlike the results from the HIJING model, the $D$ values for net-charge 
obtained from the HRG calculations are higher compared to net-baryon, with or 
without the resonance decay contributions. This may be due to the fact that 
charged particles with higher electric charge contribute to the higher 
net-charge fluctuations in HRG.

\bef[ht!]
\includegraphics[width=0.5\textwidth]{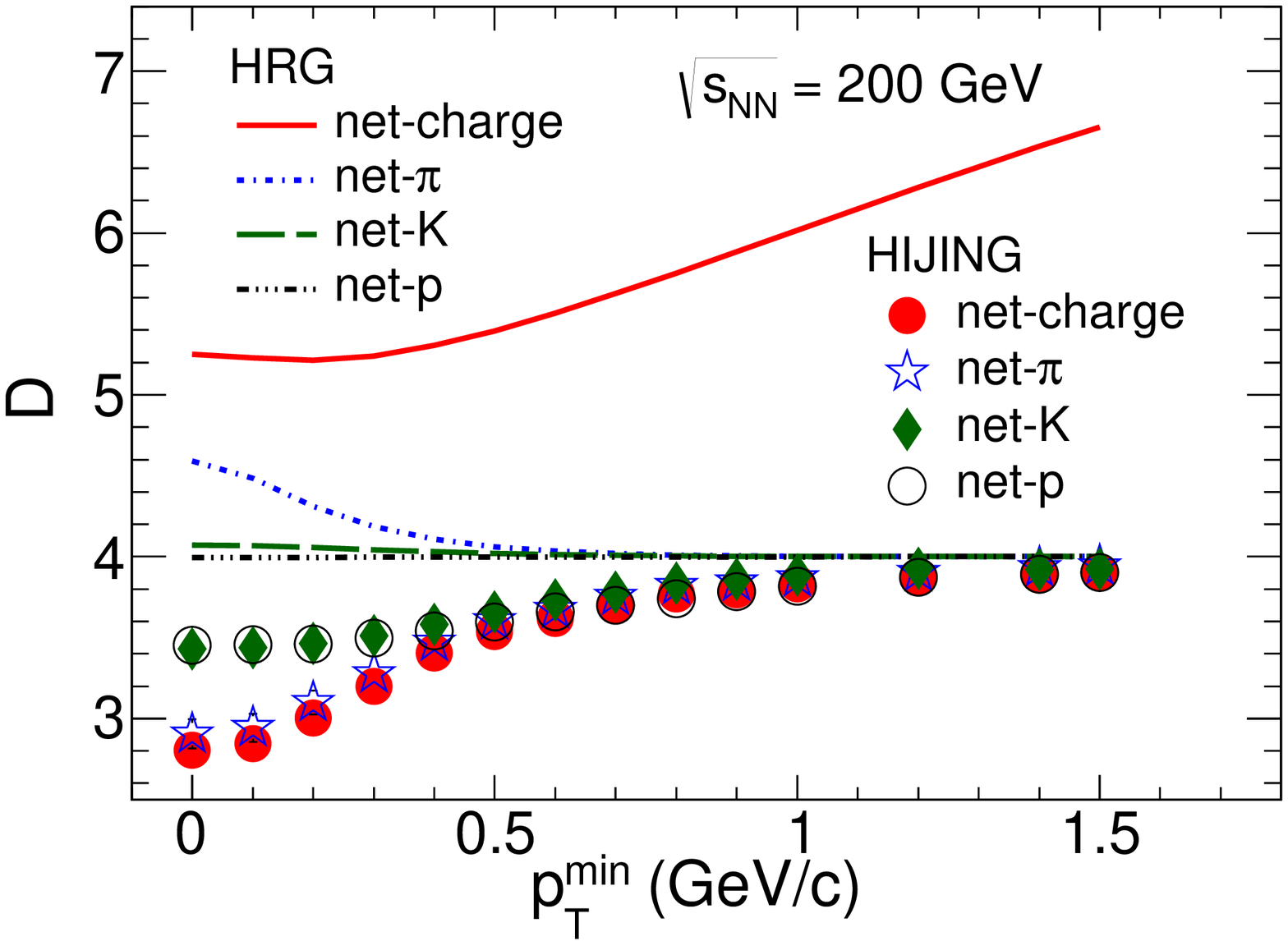}
\caption{The lower $p_T$ selection range dependence of the $D$ for net-charge, 
net-pion, net-kaon, and net-proton fluctuations calculated for (0\%--5\%) 
centrality in Au$+$Au collisions at \sqsn = 200 GeV with the HIJING model. The 
$D$ from HRG calculations is compared with HIJING model calculations.}
\label{fig:lowptcutVsD}
\eef

\subsection{Lower $p_T$ selection threshold dependence of $D$-measure}
A different experiment might have different lower $p_{T}$ acceptance for 
particles, hence it is important to study the effect of lower $p_{T}$ selection 
threshold on the selected charged particles. Figure~\ref{fig:lowptcutVsD} shows 
the $p_T^{min}$ selection threshold dependence of $D$ for net-charge, net-pion, 
net-kaon, and net-proton fluctuations in Au$+$Au collisions at \sqsn = 200 GeV 
using the HIJING model. The considered particles are within $|\eta|<$ 0.5 and 
maximum $p_T$ is 5.0 GeV/$c$. Since the net-charge fluctuations are dominated by 
net-pion fluctuations, the net-charge and net-pion fluctuations are similar and 
increase with $p_T^{min}$ selection threshold. A saturation trend is observed 
in $D$ after $p_T^{\mathrm {min}} \gtrsim$ 0.5 GeV/$c$. The lower $p_T$ 
selection threshold has a smaller effect on net-kaon and net-proton fluctuation as 
compared to net-pion fluctuations. In case of kaons and protons, the mass is 
already so large that the quantum statistics effect is almost negligible, and 
thus the 
result can be regarded as that obtained with Boltzmann approximation. The 
results from the HIJING model are compared with the $D$ values from the HRG 
calculations. The net-kaon and net-proton results from the HRG calculations are 
least affected by the $p_T^{min}$ selection threshold. 
The net-pion results are affected at lower $p_T$ selection threshold due to the 
quantum statistics and at higher $p_T^{min}$ selection threshold, the pion 
momentum distribution can be approximated with Boltzmann statistics. The 
increasing trend in net-charge fluctuation in HRG calculation is due to the 
contribution from the resonances with higher electric charge states such as 
$\Delta^{++}$.

The evolution of net-charge fluctuations with collision energy has been studied 
by various experiments at SPS~\cite{Sako:2004pw}, RHIC~\cite{Abelev:2008jg}, and 
LHC~\cite{Abelev:2012pv} energies. Figure~\ref{fig:eneVsD} shows the collision 
energy dependence of $D$ for net-charge fluctuations in most central 
(0\%--5\%) collisions. The experimental data shows a systematic decrease of 
$D$ value as one goes from lower to higher collision energies. The 
experimental measurements are compared with model calculations for net-charge 
fluctuations. The model calculations are performed within the same acceptance 
as experimental acceptances. The results from both HIJING and UrQMD models agree 
with each other and are independent of collision energies. The HRG calculations 
for net-charge fluctuations with and without inclusion of resonance decay are 
also shown in Fig.~\ref{fig:eneVsD}. The $D$ values of net-charge from 
HRG calculations decrease at lower \sqsn and remains constant at higher 
energies. By including the resonance decay contributions, the $D$ values are 
independent of \sqsn, which can be considered as a baseline for comparison with 
the experimental data. The experimentally measured $D$ values at lower 
energies are closer to the HRG calculation with inclusion of resonance decay 
contributions and deviates from the HRG calculation for higher energies. In 
Ref.~\cite{Jeon:1999gr}, it is shown that the $D$ value for hadron gas 
with resonance decay contributions is $\simeq$ 3 and for the QGP phase is $\simeq$ 
1.0$-$1.5. The ALICE experiment reported the $D$ value of 2.3 $\pm$ 0.22 for 
$\Delta\eta =$ 1.6 at \sqsn = 2.76 TeV~\cite{Abelev:2012pv}. This value is lower 
than the results at lower energies and in-between hadron gas and QGP 
prediction~\cite{Jeon:1999gr}. In the present study with the HRG model, we 
estimate the $D$ value $\simeq$ 4.4 by taking proper kinematic acceptance and 
resonance decay contributions. This serves as a more realistic baseline for 
comparison of experimental data in the hadronic phase. Comparing the 
experimental results from Ref.~\cite{Abelev:2012pv} with the HRG calculation, it 
can be inferred that the experimentally measured $D$ value at LHC energy is 
substantially lower than the HRG calculation and closer to those for the 
expectations in the QGP phase.

\bef[t]
\bc
\includegraphics[width=0.5\textwidth]{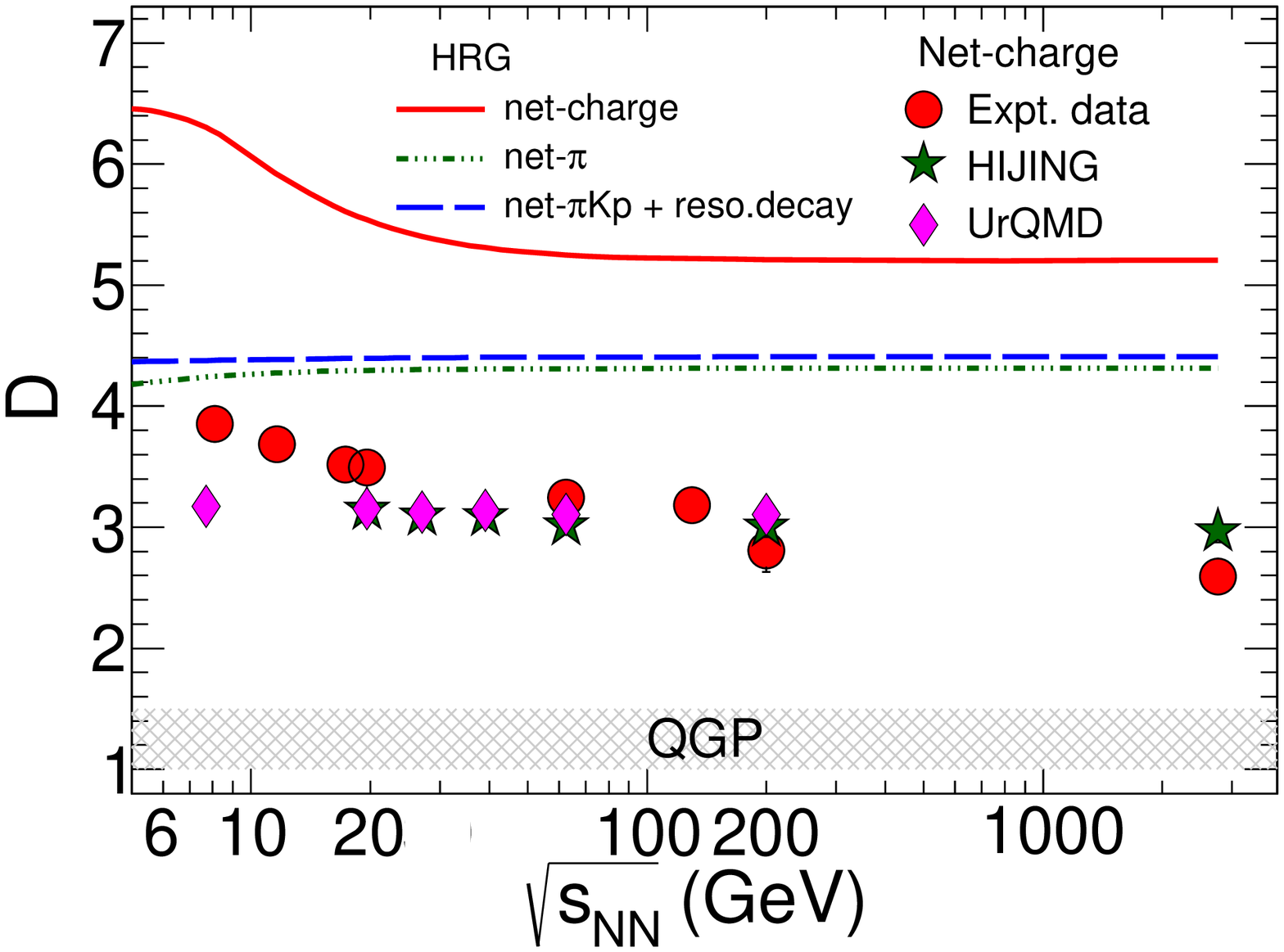}
\caption{Collision energy dependence of $D$ for the net-charge 
fluctuations measured at mid-rapidity in central (0\%--5\%) heavy-ion collisions 
at SPS~\cite{Sako:2004pw}, RHIC~\cite{Abelev:2008jg}, and 
LHC~\cite{Abelev:2012pv} energies. The model calculations (HIJING, UrQMD and 
HRG) for different energies are also shown.}
\label{fig:eneVsD}
\ec
\eef

\section{Summary}
\label{sec:summary}
In summary, we have studied net-charge fluctuation using the $D$-measure
observable within the ambit of HIJING, UrQMD and HRG models. The $D$ values are 
estimated up to higher $\Delta\eta$ intervals. A stronger dependence of $D$ 
value is observed at lower $\Delta\eta$ intervals and a decreasing trend 
continues up to higher $\Delta\eta$ intervals. Results from both the HIJING and 
UrQMD models agrees with each other up to \sqsn = 200 GeV. The HRG calculations 
with and without resonance decay contributions are also compared. We do not observe 
$\Delta\eta$ acceptance dependence of the studied observable in the HRG 
calculation except at lower energies when resonance decay contributions are not 
considered. However, there is a significant effect of resonance decay 
contributions to the $D$ observable. We also studied the $D$ as a function of 
$\Delta\eta$ interval for net-charge and individual stable particles. The 
net-charge fluctuation is dominated by the contribution from net-pion 
fluctuation. Both net-charge and net-pion fluctuations decrease with increasing 
$\Delta\eta$ interval. The $D$ values for net-kaons and net-protons also decrease 
with increasing $\Delta\eta$ interval with less steeper slope when compared to 
net-charge and net-pion fluctuations. The effect of lower $p_T$ selection 
threshold on the $D$ observable is also studied in HIJING and HRG models. The 
net-charge and net-pion fluctuations are more affected by $p_T^{min}$ selection 
threshold and increase with $p_T$ cutoff. The net-kaon and net-proton results 
are least affected by the $p_T$ cutoff both in HIJING and HRG calculations. The 
dissipation of signal during the evolution of the fireball from the 
hadronization to freeze-out has been estimated for different energies by fitting the
$D$-measure as a function of the $\Delta\eta$ interval with the error function. The 
extrapolation of the fitted curve does not explain the $D$ values calculated at 
higher $\Delta\eta$ intervals from HIJING and UrQMD models. It is to be noted 
that we observe more discrepancy at lower \sqsn as compared to LHC energy. 
We have also studied the $D$ for different collision energies for most central 
(0\%--5\%) collisions. The results obtained from the model calculations are 
independent of \sqsn. The experimental measurement at \sqsn = 2760 GeV is 
significantly lower than the HRG calculation and closer to QGP prediction. This 
study provides a more realistic baseline for comparison of experimental data and will 
be useful for other upcoming experiments also.

\end{document}